\documentclass[aps,preprint]{revtex4}
\usepackage{amssymb}
\usepackage{amsmath}
\usepackage{graphicx}

\begin{document}

\title{Quantum tunneling of magnetization in dipolar spin-1 condensates
under external fields}
\author{Limin Yang$^{1}$}
\author{Yunbo Zhang$^{1,2}$}
\affiliation{$^{1}$Institute of Theoretical Physics, Shanxi University, Taiyuan 030006,
P. R. China \\
$^{2}$Laboratory of Optics and Spectroscopy, Department of Physics,
University of Turku, 20014 Turku, Finland}

\begin{abstract}
We study the macroscopic quantum tunneling of magnetization of the $F=1$
spinor condensate interacting through dipole-dipole interaction with an
external magnetic field applied along the longitudinal or transverse
direction. We show that the ground state energy and the effective magnetic
moment of the system exhibit an interesting macroscopic quantum oscillation
phenomenon originating from the oscillating dependence of thermodynamic
properties of the system on the vacuum angle. Tunneling between two
degenerate minima are analyzed by means of an effective potential method and
the periodic instanton method.
\end{abstract}

\date{\today}
\maketitle

\section{Introduction}

Rapid experimental progresses in realization of spinor condensates \cite%
{Ho,Stamper-Kurn} have generated fascinating opportunities to study the spin
dynamics and magnetic properties of condensate atoms. The properties of
spinor condensates under external magnetic field are investigated both
experimentally \cite{Schmaljohann} and theoretically \cite{Pu1}. The
spin-exchange interaction plays an important role in these works, which is
reminiscent of the exchange interaction responsible for interesting magnetic
properties in solid. Other interactions, due to much weaker than the
exchange interaction, are in most cases ignored.

Since the spin degree of freedom becomes accessible in an optical trap, the
magnetic dipole-dipole interaction which arises from the intrinsic or
field-included magnetic dipole moment \cite{Yi1,Goral} should be taken into
account. The dipolar coupling was first considered between atoms at
different sites, based on the assumption that the spin-exchange interaction
dominates the on-site interaction \cite{Yi2}. Due to its long range and
vectorial characters, more and more attentions are paid to this spinor
dipolar condensate, for example, the ground state structure and spin
dynamics were examined for this novel quantum system in a single trap \cite%
{Cheng} and in deep optical lattices \cite{Pu2}.

Recently a major experimental breakthrough \cite{Griesmaier} has been
achieved in the condensation of chromium atoms $^{52}$Cr which possesses a
large magnetic dipole moment of $6\mu _{B}$ ($\mu _{B}$ is the Bohr
magneton) in its ground state. The dipolar interaction in this condensate is
a factor of 36 higher than that for alkali atoms, which makes possible the
study of many dipolar phenomena and new kinds of quantum phase transitions
predicted by theory. Indeed, the long-range and anisotropic magnetic
dipole-dipole interaction in degenerate quantum gases has been shown to lead
to an anisotropic deformation of the expanding Cr-BEC which depends on the
orientation of the atomic dipole moments \cite{Stuhler}.

In the present paper we mainly consider a rich set of macroscopic quantum
phenomena occurred in different quantum phases of spinor dipolar condensate.
These phases may be tuned via modifying the trapping geometry so that
various effective strengths of the dipolar interaction can be achieved.
Based on the ground state structure of a spin-1 condensate with
dipole-dipole interaction at zero temperature, macroscopic quantum tunneling
and oscillations occur in different phases, which are quite similar to what
happens in molecular magnets \cite{Garg,Kou}.

The paper is organized as follows. In section 2, we introduce the dipolar
spinor BEC model for $T=0$ under the single mode approximation, taking as a
starting point the spin system analogous to the magnetization tunneling in
magnetic particles. The ground state structure of the system is summarized
in Section 3 for zero, longitudinal and transverse fields, respectively.
Sections 4 and 5 are devoted to analyze the macroscopic quantum phenomena in
the ground state for different phases under longitudinal or transversal
external fields, in which cases tunneling between magnetizations arise
naturally. Finally a brief summary is given in Section 6.

\section{Spinor Condensate with Dipolar Interaction}

Our starting point is the many-body Hamiltonian $H$ proposed in Ref. \cite%
{Yi2}, which describes a $F=1$ spinor condensate at zero temperature trapped
in an axially symmetric potential $V_{ext}$. Without loss of generality, the
symmetry axis is conveniently chosen to be the quantization axis $\widehat{z}
$. We consider here two atomic interaction terms, the short-range
collisional interaction and the long-range magnetic dipolar interaction, the
competition of which gives rise to different quantum phases. Under an
external magnetic field $B$, the second quantized Hamiltonian of the system
reads
\begin{align}
H& =\int d\mathbf{r}\widehat{\psi }_{\alpha }^{\dagger }(\mathbf{r})[(-\frac{%
\hbar ^{2}\nabla ^{2}}{2M}+V_{ext})\delta _{\alpha \beta }-g\mu _{B}\mathbf{B%
}\cdot \mathbf{F}_{\alpha \beta }]\widehat{\psi }_{\beta }(\mathbf{r})
\notag \\
& +\frac{c_{0}}{2}\int d\mathbf{r}\widehat{\psi }_{\alpha }^{\dagger }(%
\mathbf{r})\widehat{\psi }_{\beta }^{\dagger }(\mathbf{r})\widehat{\psi }%
_{\alpha }(\mathbf{r})\widehat{\psi }_{\beta }(\mathbf{r})  \notag \\
& +\frac{c_{2}}{2}\int d\mathbf{r}\widehat{\psi }_{\alpha }^{\dagger }(%
\mathbf{r})\widehat{\psi }_{\delta }^{\dagger }(\mathbf{r})\mathbf{F}%
_{\alpha \beta }\cdot \mathbf{F}_{\delta \gamma }\widehat{\psi }_{\beta }(%
\mathbf{r})\widehat{\psi }_{\gamma }(\mathbf{r})  \notag \\
& +\frac{c_{d}}{2}\int \int \frac{d\mathbf{r}d\mathbf{r}^{\prime} }{|\mathbf{r}-%
\mathbf{r}^{\prime} |^{3}}\left[ \widehat{\psi }_{\alpha }^{\dagger }(\mathbf{r}%
)\widehat{\psi }_{\delta }^{\dagger }(\mathbf{r}^{\prime}
)\mathbf{F}_{\alpha
\beta }\cdot \mathbf{F}_{\delta \gamma }\widehat{\psi }_{\beta }(\mathbf{r})%
\widehat{\psi }_{\gamma }(\mathbf{r}^{\prime} )\right.  \notag \\
& \left. -3\widehat{\psi }_{\alpha }^{\dagger }(\mathbf{r})\widehat{\psi }%
_{\delta }^{\dagger }(\mathbf{r}^{\prime })\left( \mathbf{F}_{\alpha \beta
}\cdot \mathbf{e}\right) \left( \mathbf{F}_{\delta \gamma }\cdot \mathbf{e}%
\right) \widehat{\psi }_{\beta }(\mathbf{r})\widehat{\psi }_{\gamma }(%
\mathbf{r}^{\prime })\right] ,  \label{1}
\end{align}%
where $\widehat{\psi }_{\alpha }\left( \mathbf{r}\right) \left( \alpha
=0,\pm 1\right) $ are the field annihilation operators for an atom in the
hyperfine state $|F=1,m_{F}=\alpha \rangle .$ The two coefficients $%
c_{0}=4\pi \hbar ^{2}\left( a_{0}+2a_{2}\right) /3M$ and $c_{2}=4\pi \hbar
^{2}\left( a_{2}-a_{0}\right) /3M$ characterize the density-density and
spin-spin collisional interactions, respectively. Here $a_{f}$ $(f=0$ or $2)$
being the $s$-wave scattering length for spin-1 atoms in the combined
symmetric channel of total spin $f$. The dipolar interaction parameter is $%
c_{d}=\mu _{0}g^{2}\mu _{B}^{2}/4\pi $ with $g$ being Land\'{e}
$g$-factor. The $V_{ext}$ represent the external trapping potential
which is spin independent for a far off-resonant optical trap. And
$\mathbf{e}=\left( \mathbf{r-r}^{\prime }\right)
/|\mathbf{r-r}^{\prime }|$ is a unit vector. In this study, the
external field $\mathbf{B}$ is assumed to be spatially uniform.

In order to simplify the Hamiltonian (\ref{1}), we usually adopt the
single mode approximation (SMA): $\widehat{\psi }_{\alpha
}(\mathbf{r})\simeq \phi \left( \mathbf{r}\right)
\widehat{a}_{\alpha }$ \cite{Yi3}, where $\phi \left(
\mathbf{r}\right) $ is the spin independent spatial wave function of
the condensate, $\widehat{a}_{\alpha }$ is the annihilation operator
for $ m_{F}=\alpha $ component. It is always safe to use this
approximation for ferromagnetic interactions, however, it may break
down for antiferromagnetic interactions if both the atomic number
$N$ and the magnetization $\mathcal{M} $ are large. The interaction
parameters in our case of study are such that $ \left\vert
c_{2}\right\vert \ll c_{0}$ and $c_{d}\lesssim 0.1\left\vert
c_{2}\right\vert $. Under these conditions, the single mode
approximation is expected to be valid if the trapping potential is
axially symmetric. All integral terms in the long range interactions
involving $e^{\pm i\varphi _{e}}$ vanish after integrating over the
azimuthal angles of $\mathbf{r-r} ^{\prime }$. We then obtain a much
simpler Hamiltonian by assuming that the mode function $\phi \left(
\mathbf{r}\right) $ hence possesses the axial symmetry and dropping
the spin-independent constant terms \cite{Yi2}:
\begin{equation}
H=\left( c_{2}^{\prime }-c_{d}^{\prime }\right) \mathbf{\hat{L}}%
^{2}+3c_{d}^{\prime }\left( \hat{L}_{z}^{2}+\widehat{n}_{0}\right) -g\mu _{B}%
\mathbf{B}\cdot \mathbf{\hat{L},}  \label{2}
\end{equation}%
where $\mathbf{\hat{L}}=\widehat{a}_{\alpha }^{\dagger }\mathbf{F}_{\alpha
\beta }\widehat{a}_{\beta }$ characterizes the total many-body angular
momentum operator, $\widehat{n}_{0}=\widehat{a}_{0}^{\dagger }\widehat{a}%
_{0} $ is the number operator for $m_{F}=0$ atoms. The new parameters are $%
c_{2}^{\prime }=\left( c_{2}/2\right) \int d\mathbf{r}|\phi \left( \mathbf{r}%
\right) |^{4}$ and $c_{d}^{\prime }=\left( c_{d}/4\right) \int d\mathbf{r}d%
\mathbf{r}^{\prime }|\phi \left( \mathbf{r}\right) |^{2}|\phi \left( \mathbf{%
r}^{\prime }\right) |^{2}\left( 1-3\cos ^{2}\theta _{e}\right) /|\mathbf{r}-%
\mathbf{r}^{\prime }|^{3}$, with $\theta _{e}$ being the polar angle of $%
\left( \mathbf{r}-\mathbf{r}^{\prime }\right) $. Eq. (\ref{2}) may be put
into the following dimensionless form by rescaling it in energy unit $%
|c_{2}^{\prime }|$:%
\begin{equation}
H=\left( \pm 1-c\right) \mathbf{\hat{L}}^{2}+3c\left( \hat{L}_{z}^{2}+%
\widehat{n}_{0}\right) -\mathbf{B}^{\prime }\cdot \mathbf{\hat{L},}
\label{3}
\end{equation}%
where $+$ and $-$ correspond to $c_{2}>0$ and $c_{2}<0$, respectively. The
new parameter $c=c_{d}^{\prime }/|c_{2}^{\prime }|$ thus measures the
relative strength of dipolar interaction with respect to the spin-exchange
interaction. The dimensionless magnetic field $\mathbf{B}^{\prime }=g\mu _{B}%
\mathbf{B}/\left\vert c_{2}^{\prime }\right\vert =B^{\prime }(\sin \theta
,0,\cos \theta )$ is assumed to lie in the $xz$-plane with an angle $\theta $
relative to the $z$-axis.

\section{Ground State Structure under External Field}

We summarize the ground state structure which has been described in Ref.
\cite{Yi2}. In the absence of an external field, the ground state of our
system is divided into three distinct regions $A$, $B$, and $C$ in the $%
c_{2}^{\prime }$-$c_{d}^{\prime }$ parameter plane (see Figure 1). We denote
here the simultaneous eigenstate of $\mathbf{\hat{L}}^{2}$ and $\hat{L}_{z}$
as $\left\vert l,m\right\rangle $ with eigenvalues $l(l+1)\hbar ^{2}$ and $%
m\hbar $, respectively. In Region $A$ ($c_{d}^{\prime }>0$ and $%
c_{d}^{\prime }>c_{2}^{\prime }$) the ground state is given by $G=\left\vert
N,0\right\rangle $, a quantum superposition of a chain of Fock states $%
\left\vert N/2-k,2k,N/2-k\right\rangle $ in which the numbers of atoms in
the spins $1$ and $-1$ are equal. In Region $B$ ($c_{d}^{\prime }<0$ and $%
c_{d}^{\prime }<-c_{2}^{\prime }/2$), $G=\left\vert N,\pm N\right\rangle $
is a Fock state with all the population in either $m_{F}=1$ or $-1$. In
these two regions, the $\widehat{n}_{0}$ term is at least a factor of $1/N$
smaller than the rest and therefore can be neglected safely. In Region $C$,
however, the ground state is a little more complicated because the $\widehat{%
n}_{0}$ term is expected to be important. In general it is expressed as $%
G=\sum_{l}g_{l}\left\vert l,0\right\rangle $, a superposition of different
angular momentum states with $\left\langle \hat{L}_{z}\right\rangle =0$.

When we apply a longitudinal field to the system along the $z$-axis ($\theta
=0$), the condensate in different parameter regions behave quite
differently. In region $A$, the system can be mapped onto an easy-plane
anisotropic particle with the transverse $xy$-plane being the easy-plane.
The ground state is $G=\left\vert N,m\right\rangle $ with $m$ depends on the
field strength linearly, with steps. Region $B$ corresponds to a uniaxial
anisotropic magnetic spin model with the easy axis $z$. The presence of the
external magnetic field simply removes the two-fold degeneracy and forces
the atoms into the fully polarized state $G=\left\vert N,N\right\rangle $.
The ground state in Region $C$ is changed into $G=\sum_{l\geqslant
m_{0}}g_{l}\left\vert l,m_{0}\right\rangle $ while $m_{0}$ increases with
the field strength.

Now we check the situation when a transverse field is applied along the $x$%
-axis, i.e., $\theta =\pi /2$. Due to the easy plane anisotropy in Region $A$%
, the ground state of the condensate is fully polarized, however, in this
case along the $x$-axis, $G=\left\vert N,0\right\rangle _{m_{x}=N}$. The
situation in Region $B$ is more interesting because the ground state here $%
G=\sum_{m}g_{m}\left\vert N,m\right\rangle $ is two-fold degenerate and it
provides another example that can exhibit macroscopic quantum tunneling of
magnetization. Stepwise magnetization curve will appear in Region $C$ - each
step means the breaking of one spin singlet pair.

For clarity, we choose two of above models which admit extensive study of
tunneling of magnetization, i.e., Region $A$ with a longitudinal field and
Region $B$ with a transverse field. The dipolar spinor condensate thus
provides another platform for the investigation of macroscopic quantum
phenomena.

\section{Macroscopic Quantum Oscillation of Magnetization}

We first consider the spinor dipolar condensate under longitudinal field
along $z$-axis in Region $A$, in which case the transverse $xy$-plane
corresponds to an easy-plane anisotropy. After dropping the unimportant $%
\widehat{n}_{0}$ and constant $\mathbf{\hat{L}}^{2}$ terms, the effective
Hamiltonian can be obtained%
\begin{equation}
H_{LA}=3c\hat{L}_{z}^{2}-B^{\prime }\hat{L}_{z},
\end{equation}
where $c>0$. The model is precisely the same as that for a ferromagnetic
particle \cite{Garg,Kou} with easy-plane anisotropy and a magnetic field
along hard axis. The Hamiltonian is exactly diagonal in terms of the
eigenstate $\widehat{L}_{z}$ and with the eigenvalue $E_{m}=3cm^{2}-B^{%
\prime }m$. For zero temperature the magnetization increases
stepwise as a consequence of the fact that $m$ take only integer
values. This model, on the other hand, provides a perfect
manifestation of the $\Theta $ vacuum effect originating from the
oscillating dependence of thermodynamic properties of the system on
the ``vacuum angle" \cite{KriveRoz}. The concept of the $\Theta $
vacuum was developed mainly for the models of modern quantum field
theory \cite{Rajaraman}, but also in condensed media the
nonperturbative vacuum is not a mathematical abstraction. The vacuum
angle is nothing but the factor with which the total time derivative
enters the Lagrangian. For Aharonov-Bohm problem in conductors with
charge density waves, $\Theta $ is the normalized magnetic flux
\cite{Bogachek}. In the Josephson junction of mesoscopic sizes, the
vacuum angle depends on the voltage applied to the junction
\cite{Krive}. We see shortly the magnetic field enters the
Lagrangian and plays the role of vacuum angle in our spinor dipolar
condensate system.

We express the partition function as a spin coherent state path integral for
large number of atoms $N\gg 1$%
\begin{equation}
Z=Tr\exp \left( -\beta H_{LA}\right) =\int D\left\{ \mu \left( \mathbf{n}%
\right) \right\} \exp \left( -S_{E}\right) ,
\end{equation}%
where the measure of the integration is decomposed into $D\left\{
\mu \left( \mathbf{n}\right) \right\} =\prod_{k=1}^{N-1}\left( N\sin
\theta _{k}d\phi _{k}d\theta _{k}/2\pi \right) $. The semiclassical
approximation of the partition function turns out to be the
transition amplitude between two spin configurations conneted by
periodic orbits with fixed imaginary time period $ \beta $.
Following the usual procedure \cite{Pereromov}, we represent the the
state vector of the system as coherent states and slice the integral
into $\mathcal{N}$ identical pieces of length $\epsilon =\beta
/\mathcal{N}$ . Inserting complete sets of states gives
\begin{equation}
Z=\left\langle \mathbf{n}_{F}\right\vert \exp \left( -\beta
H_{LA}\right) \left\vert \mathbf{n}_{I}\right\rangle
=\prod\limits_{k=1}^{\mathcal{N} -1}\int D\left\{ \mu _{k}\left(
\mathbf{n}\right) \right\}
\prod\limits_{k=0}^{\mathcal{N}-1}\left\langle
\mathbf{n}_{k+1}\right\vert (1-\epsilon H_{LA})\left\vert
\mathbf{n}_{k}\right\rangle
\end{equation}%
After some algebraic evaluation and finally passing to the time
continuum limit $\mathcal{N}\longrightarrow \infty $ we obtain the
Euclidean action in
imaginary time $\tau =it$ as (dots now denote $\tau $-derivatives)%
\begin{equation}
S_{E}=\int_{0}^{\beta \hbar }d\tau \lbrack iN\dot{\phi}(1-\cos \theta
)+3cN(N+1)\cos ^{2}\theta -B^{\prime }N\cos \theta ]
\end{equation}%
with $\beta =1/k_{B}T$ and $T$ the temperature. Integrating over $\cos
\theta $ we map the magnetic system onto a particle problem with Lagrangian%
\begin{equation}
\mathcal{L}=\frac{m_{eff}\dot{\phi}^{2}}{2}+i\Theta
\dot{\phi},\label{lag}
\end{equation}%
where the effective mass $m_{eff}=1/6c$ and $\Theta =N\left( 1-B^{\prime
}/6cN\right) $. This Lagrangian is exactly the one for $\Theta $ vacuum of
the non-Abelian gauge field in references \cite{Krive,Rajaraman}. We noticed
that the second term of $\mathcal{L}$ is the total imaginary time derivative
and has no effect on the classical equation of motion, while it indeed
alerts the canonical momentum into $\Pi _{\phi }=m_{eff}\dot{\phi}+i\Theta $%
. In order to minimize the Euclidean action $S_{E}$, we try to find the
classical configuration, that is, the periodic instanton solution under the
boundary condition $\phi _{n}\left( \tau +\beta \right) =\phi _{n}+2\pi n$.
The result is $\phi _{n}=2\pi n\tau /\beta $ and the Euclidean action for
this solution is $S_{E}=s_{0}n^{2}+i2n\pi \Theta $, where $n$ is the winding
number characterizing homotopically nonequivalent classes and $s_{0}=\pi
^{2}k_{B}T/3c.$ The Euclidean functional integral of the partition function
contains thus an additional summation over the homotopic number and detailed
calculation leads to%
\begin{equation}
Z=\sum_{n=-\infty }^{\infty }Z_{n}=\vartheta _{3}\left[ \Theta ,\exp (-s_{0})%
\right] ,
\end{equation}%
where $\vartheta _{3}\left( v,q\right) $ is the Jacobi theta function
oscillating with $\Theta $. By means of the well known asymptotics of the
Jacobi theta function, the ground state energy can be shown clearly
oscillating with $\Theta $, i.e. the external magnetic field $B$
\begin{equation}
E_{0}=-k_{B}T\ln Z=-\frac{(B^{\prime })^{2}}{12c}+\frac{1}{2m_{eff}}%
\{\{\Theta \}\}^{2},
\end{equation}%
where $\{\{x\}\}$ is the difference between $x$ and its nearest integer.
According to this, the external magnetic field induces quantum oscillations
in the dipolar spinor condensate. The period of oscillation is shown to be $%
\delta B=6c_{d}^{\prime }/g\mu _{B}$ (see Figure 2).

The topological term in the Lagrangian (\ref{lag}) leads to the
oscillation behavior of our system. We emphasize here the ground
state (or in the language of quantum field theory, the instanton
vacuum) of the condensate acquires a vacuum angle owing to the
presence of the external magnetic field, which breaks the symmetry
and changes the topology of the system. In a charge density wave
ring-shaped conductor placed in an external vector potential field
the magnetic susceptibility and the electrical conductivity
oscillate as a function of the flux \cite{Bogachek}, while the
voltage applied on the Josephson junction induces the oscillation of
the effective capacitance of the junction \cite{Krive}. In our
system of investigation the macroscopic quantum effect manifests
itself by the oscillation of the effective magnetic momentum
$M=-k_{B}T\partial \ln Z/\partial B$ as a function of the applied
magnetic field. The magnetization curve at zero temperature is
stepwise due to the macroscopic quantum oscillation of the effective
magnetic moment and the period of oscillation depends on the
strength of the dipolar interaction $c_{d}^{\prime }$ uniquely. We
show this oscillation period in Figure 2, together with the
mathematical function $\{\{x\}\}$. The functional integral approach
presented here becomes necessary when environmental friction or
dissipation is included, which amounts to the introduction into the
Euclidean Lagrangian of an additional term describing the non-local
interaction of the instantons \cite{CL}. This is, however, beyond
the scope of this paper. It is interesting to estimate the
modulation period in the typical laboratory experiment on spinor
condensates. For a condensate with $N\sim 5\times 10^{5}$ sodium
atoms $^{23}$Na and density $\rho \sim 10^{14}cm^{-3}$ \cite{Davis}
we obtain the regular interval of magnetic field $\delta B\simeq
1.4Gauss$. This oscillation period can be as small as $1.5\times
10^{-3}Gauss$ for $ 2\times 10^{4}$ rubidium atoms $^{87}$Rb with
density $2.6\times 10^{12}cm^{-3}$ as in the first experimental
achievement of BEC in JILA \cite {Anderson} and can be as large as
$10.5Gauss$ for $5\times 10^{4}$ chromium atoms $^{52}$Cr with
density $10^{14}cm^{-3}$ as in the latest dipolar condensate
experiments in Stuttgart \cite{Griesmaier,Stuhler}. Moreover, it can
be adjusted flexibly by changing the trapping potential geometry.

\section{Quantum Tunneling of Magnetization between two Local Minima}

Now we prepare the system properly in Region $B$ and apply a transverse
field along the $x$-axis to the condensate. Again we drop the constant terms
and are left with the following effective Hamiltonian%
\begin{equation}
H_{TB}=-3d\widehat{L}_{z}^{2}-B^{\prime }\hat{L}_{x}
\end{equation}%
where $d=\left\vert c\right\vert $. The model describes a quantum spin
system with the easy-axis anisotropy while the external field is along $x$%
-axis. This model has been extensively studied in the context of spin
tunneling. Classically under the influence of a weak transverse field along
the $x$ direction, the two energy minima move away from the zero filed
positions ($+z$ or $-z$) and towards $x$-axis while remaining in the $xz$
plane. For $0\leq B^{\prime }\leq 6dN$, they are located at $\gamma
_{-}=\arcsin \left( B^{\prime }/6dN\right) $ and $\gamma _{+}=\pi -\theta
_{-}$, respectively, with the angle $\gamma $ between $\mathbf{\hat{L}}$ and
$z$-axis. The degeneracy is removed when $B^{\prime }\geq B_{sat}^{\prime
}=6dN$, where the system is completely polarized by the external field and
the two minima merge along $x$-axis. Quantum mechanically, the degeneracy is
lifted before the magnetic field reaches $B_{sat}^{\prime }$ due to the
magnetization tunneling. A well-known consequence of the tunneling between
two degenerate states is the lifting of their degeneracy: The two new
eigenstates are a symmetric and an antisymmetric superposition of the
original states characterized by an energy difference (or tunneling
splitting) $\Delta E_{0}$ inversely proportional to the tunneling rate. The
quantity of interest to determine the occurrence of tunneling is therefore
this energy difference between the two lowest eigenstates of the Hamiltonian.

To calculate analytically this energy splitting, we use the effective
potential method \cite{Zaslavskii}\ which maps the spin system onto a
particle system and the result can be easily obtained with the periodic
instanton method. The Schr\"{o}dinger equation $H_{TB}\left\vert \Psi
\right\rangle =E\left\vert \Psi \right\rangle $ in the $\hat{L}_{z}$
representation takes the form
\begin{eqnarray}
2\left( E+3dm^{2}\right) C_{m}+B^{\prime }\sqrt{\left( N-m\right) \left(
N+m+1\right) }C_{m+1} &&  \notag \\
+B^{\prime }\sqrt{\left( N+m\right) \left( N-m+1\right) }C_{m-1}=0 &&
\label{she}
\end{eqnarray}%
where $m=-N,-N+1,\cdots N$ and $C_{m}=0$ for $|m|>N$.

Let us introduce the generating function,
\begin{equation}
\Phi =\sum_{m=-N}^{N}\frac{C_{m}}{\sqrt{\left( N-m\right) !\left( N+m\right)
!}}\exp \left( mx\right) .
\end{equation}%
Multiplying equation (\ref{she}) by a factor $\exp \left( mx\right) /\sqrt{%
\left( N-m\right) !\left( N+m\right) !}$ and summing all terms with $-N\leq
m\leq N$, we can transform the equation (\ref{she}) into
\begin{equation}
3d\Phi ^{\prime \prime }-B^{\prime }\sinh x\Phi ^{\prime }+\left(
E+B^{\prime }N\cosh x\right) \Phi =0.
\end{equation}

In order to remove the first derivative term, let us define a new function $%
\Psi =\Phi \exp (-\frac{1}{2}B^{\prime }\cosh x)$. When we replace $\Phi $
with $\Psi $, the new Schr\"{o}dinger equation can be written after dividing
by $N^{2}$ in the form%
\begin{equation}
N^{-2}\Psi ^{\prime \prime }+\Psi \left( \kappa -U\right) =0,
\end{equation}%
where the corresponding parameter $\kappa $ describe the dimensionless
energy $\kappa =E/3dN^{2}$, and $U=(a\cosh x-1)^{2}-a^{2}$ is the effective
potential well with $a=B^{\prime }/6dN$.

The value $N^{-1}$ plays the role of the Planck constant $\hbar $, the
potential takes the form of a double well for $a<1$, i.e., for external
magnetic field not exceeding the saturation value $B_{sat}^{\prime }=6dN$.
The two local minima thus play the role of the degenerate classical states
and the energy splitting of the lower states takes the following form
according to \cite{Weiss}%
\begin{equation}
\Delta E_{n}=\Delta E_{0}q^{n}/n!
\end{equation}%
with the splitting for the ground state%
\begin{equation}
\Delta E_{0}=\sqrt{\hbar \omega /\pi }C\exp (-S/\hbar ).
\end{equation}%
Here $S_{E}$ is the Euclidean action evaluated along the trajectory from the
left minimum $x_{-}=-\cosh ^{-1}\left( 1/a\right) $ to the right $%
x_{+}=\cosh ^{-1}\left( 1/a\right) $, $\omega =\sqrt{1-a^{2}}$ is the small
oscillator frequency near the bottom of potential well $x_{\pm }$. The
asymptotic form of the instanton trajectory determined the constant $C$ and $%
q=C^{2}/2\hbar \omega $. For the potential $U$ we have%
\begin{equation}
S_{E}=\ln \left( \frac{1+\sqrt{1-a^{2}}}{a}\right) -\sqrt{1-a^{2}}.
\end{equation}%
Including the prefactor we have finally%
\begin{equation}
\Delta E_{0}=\frac{24d}{\sqrt{\pi }}N^{3/2}\frac{(1-a^{2})^{5/4}a^{2N}}{(1+%
\sqrt{1-a^{2}})^{2N}}\exp \left[ \left( 2S_{E}N+1\right) \sqrt{1-a^{2}}%
\right] .  \label{de}
\end{equation}

Experimentally this level splitting can be measured by means of the
resonance measurement developed by Awschalom \cite{Awschalom}. Tunneling
between two degenerate orientations of magnetization leads to the splitting
of the "non-tunneling" ground state energy level into two levels separated
by $\Delta E_{0}$. Correspondingly a very weak ac field of frequency $\Delta
E_{0}/\hbar $ will induce transitions between the two levels, which should
result in the resonant absorption of the energy of the field. The atoms in
the condensate are utterly identical and we do not have the problem of
distribution of particle sizes and shapes. The level splitting eq. (\ref{de}%
) is expressed in units of $c_{2}^{\prime }$. It is easily shown that for
the sodium condensate in \cite{Davis} the dipole-dipole interaction, i.e.,
the anisotropic energy in our model, is estimated as $c_{d}^{\prime
}=1.69\times 10^{5}Hz$ or $11.5\mu K$, a quantity much smaller than the
anisotropy energy of molecular magnets $Mn_{12}Ac$ or $Fe_{8}$ \cite{Hennion}%
. The level splitting can be greatly enhanced by a smaller number of atoms
in the condensate and also by a stronger dipolar-dipolar interaction.
Taking, as an example, sodium atoms ($N=38$) under an external field of $%
a=0.6$, we have $\Delta E_{0}^{\prime }=1.14\times 10^{2}Hz$. For the
condensate of $^{52}Cr$, the anisotropic energy is enhanced to $%
c_{d}^{\prime }=2.44\times 10^{6}Hz$ or $\allowbreak 166\mu K$. For $N=39,$
we have $\Delta E_{0}^{\prime }=3.15\times 10^{2}Hz$. The level splitting
remains in the same magnitude order because it depends very sensitively on
the total number of atoms. These data are easily accessible in the present
ultracold atom experiments.

\section{Summary}

Inspired by the macroscopic quantum tunneling in the magnetic system, we
have investigated the macroscopic quantum tunneling in the dipolar spinor
condensates at zero temperature and obtained some interesting results by
analyzing different phase areas with applied external fields. We found that
the ground state energy and the effective magnetic moment oscillate with the
external magnetic field in Region $A$ under a longitudinal field and the
oscillating period depends on the strength of the dipolar interaction as $%
\delta B=6c_{d}^{\prime }/g\mu _{B}$. This model provides a condensed media
realization of the $\Theta $ vacuum in quantum field theory. The model in
Region $B$ with a transverse field provides an example where quantum
tunneling of magnetization occurs between two local minima. We estimated the
level splitting to be at the reach of current ultracold atom experiments.

\section{Acknowledgments}

This work was supported by National Natural Science Foundation of China
(NSFC) under grant No. 90203007, Shanxi Province Youth Science Foundation
under grant No. 20051001. YZ was also partially supported by Academy of
Finland. We thank W.-D. Li, Y.-H. Nie and S.-P. Kou for useful discussions.

\bigskip \pagebreak

Figure Captions:

Figure 1: (Color online) Magnetic phase diagram of dipolar spinor
condensate parametrized in the $c_{2}^{\prime }$-$c_{d}^{\prime }$
plane. Corresponding ground states are shown for zero external
field. The two tunneling models studied in this paper are in phase
$A$ with a longitudinal field and phase $B$ with a transverse field.

Figure 2: The mathematical function $\{\{x\}\}$ and the oscillation of the
magnetization with period $\delta B$.

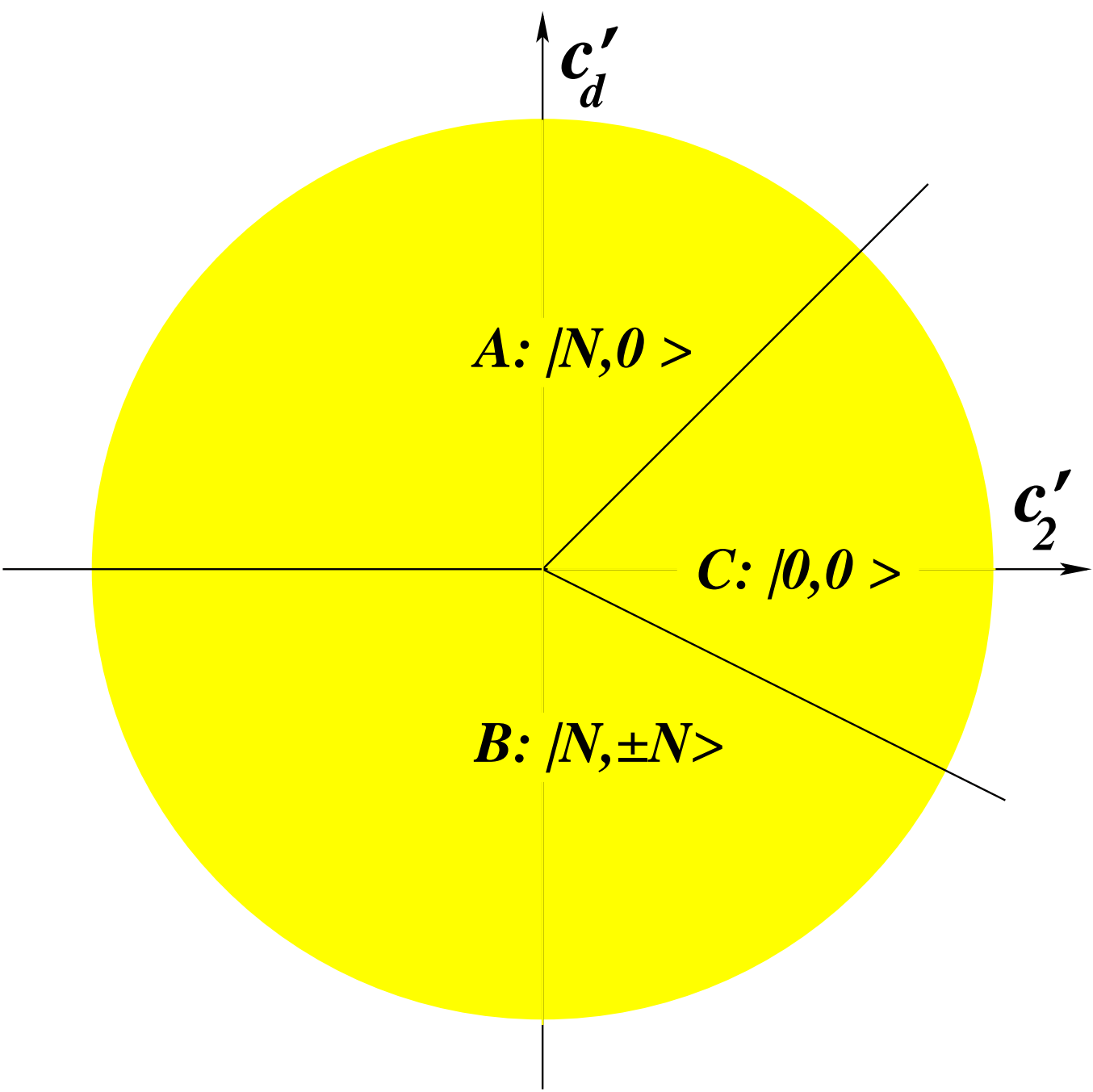
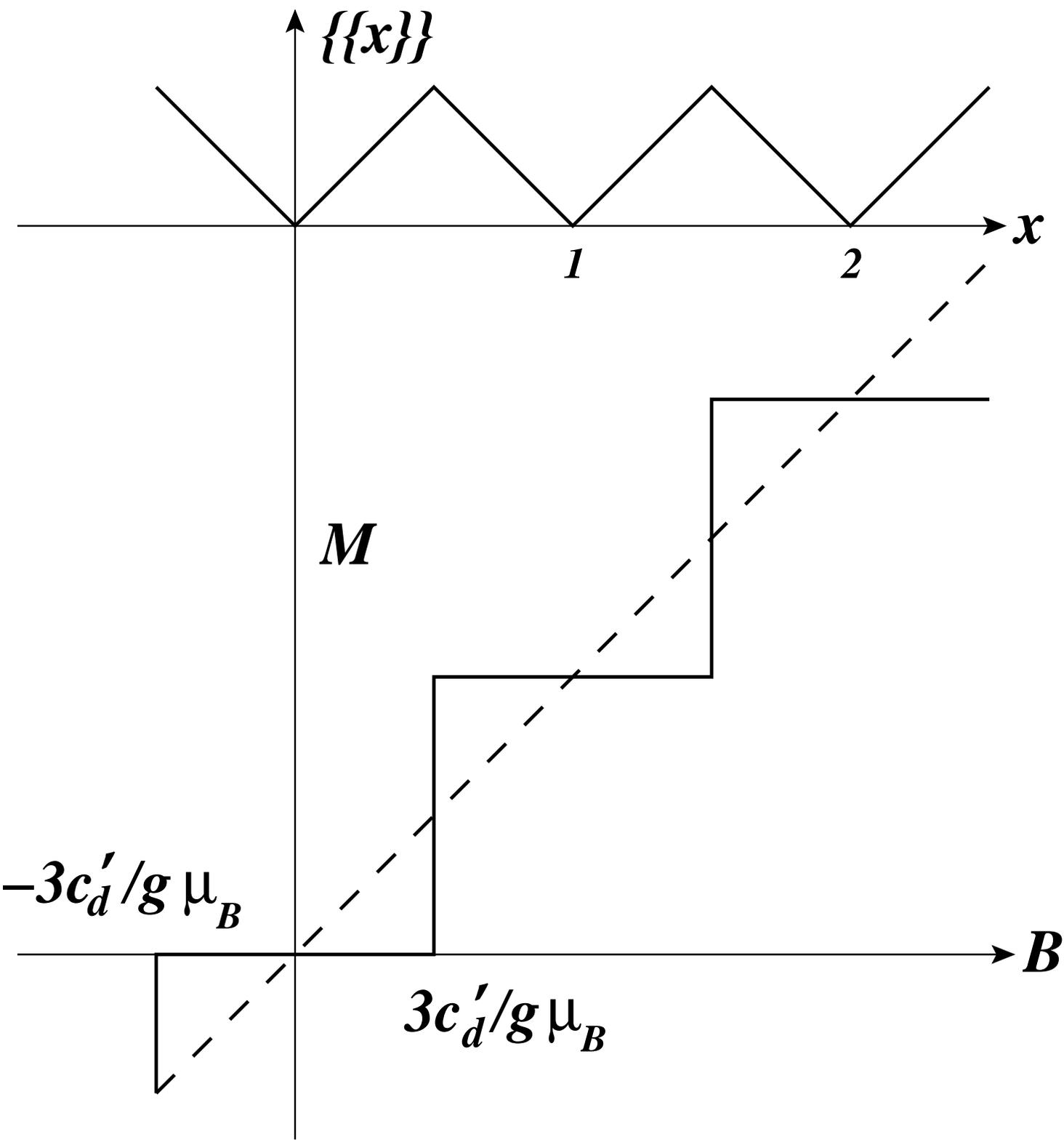
\end{document}